\def\lesssim{\mathrel{\hbox{\rlap{\hbox{\lower4pt\hbox{$\sim$}}}\hbox{$<$}}}}

\def\gtrsim{\mathrel{\hbox{\rlap{\hbox{\lower4pt\hbox{$\sim$}}}\hbox{$>$}}}}

\def\msun{M$_{\odot}$}

\def\teff{$T_{\rm eff}$}

\def\ll_lsun{log$({L/\rm L_{\odot}})$~}

\def\masa_msun{$M/ \rm M_{\odot}$~}

\def\m_mstar{$M/M_{*}$~}

\def\mct{\mbox{\object{MCT 0130$-$1937}}}
\def\hs1517{\mbox{\object{HS 1517+7403}}}

\def\gteff{$\log$ \teff$-$$\log g$}

\documentclass{aa}
\usepackage{graphicx}
        
\begin{document}

\title{Low-mass, helium-enriched PG1159 stars: a possible evolutionary
origin and the implications for their pulsational stability properties}

\author{L. G. Althaus$^{1,2}$\thanks{Member of the Carrera del Investigador
Cient\'{\i}fico y Tecnol\'ogico, CONICET, Argentina.},
A. H. C\'orsico$^{1,2\star}$
\and M. M. Miller Bertolami$^{1,2}$\thanks{Fellow of CONICET, Argentina.}}

\offprints{L. G. Althaus}

\institute{
$^1$   Facultad   de   Ciencias  Astron\'omicas   y   Geof\'{\i}sicas,
Universidad Nacional de La Plata,  Paseo del Bosque S/N, (B1900FWA) La
Plata, Argentina.\\
$^2$ Instituto  de Astrof\'{\i}sica La Plata, IALP, CONICET-UNLP\\
\email{althaus,acorsico,mmiller@fcaglp.unlp.edu.ar} }

\date{Received; accepted}

\abstract{}{ We examine a recently--proposed evolutionary scenario 
that could 
explain the existence of the low--mass, helium--enriched PG1159 stars.
We focus in particular on the study of the pulsational stability 
properties of the evolutionary models predicted by such scenario.}
{ We assess the overstability of pulsation $g-$modes of stellar models as  
evolution  proceeds along  the  PG1159 domain. Stellar models are extracted
from the full evolution of a 1-\msun\ model star that
experiences its first thermal pulse as a late thermal pulse
(LTP) after leaving the AGB.  The evolutionary stages corresponding
to the born--again episode and the subsequent helium sub-flashes are 
taken into account in detail.}{ Under reasonable mass--loss rate 
assumptions,
the evolutionary scenario reproduces the high helium abundances observed 
in some PG1159 stars. We find that, despite the high helium 
abundance in the driving 
layers, there exists  a narrow region in the \gteff\ diagram  for  which 
the  helium--enriched  PG1159  sequence
exhibits  unstable pulsation  modes with periods in the range 500 to
1600 s. In particular, the nonpulsating helium--enriched PG1159 star,
\mct, is located outside the theoretical instability domain. Our results
suggest that \mct\ is a real non-pulsating star and that 
the lack of pulsations should not be attributed to unfavorable geometry.}{
Our study 
hints at a consistent picture between the evolutionary 
scenario that could explain the existence of helium--enriched PG1159 stars 
and the nonvariable nature of \mct. We also
present theoretical support for the unusually high helium abundance
observed in the nonpulsating PG1159 star \hs1517.  We
suggest that \hs1517 could be a transition object linking the
low--mass helium--rich O(He) stars with the helium--enriched PG1159 stars
via the evolutionary connection \object{K1$-$27}\ $\rightarrow$ \hs1517\  
$\rightarrow$ \mct.}


\authorrunning{L. G. Althaus et al.}

\titlerunning{Low--mass, helium--enriched PG1159 stars.}

\maketitle


\section{Introduction}

PG1159 stars constitute the evolutionary link between the
asymptotic giant branch (AGB) stars and most of the hydrogen--deficient 
white dwarf (WD) stars. 
Currently, 37 stars are members of the PG1159 family, which span a
wide domain in the \gteff\ diagram ($g$ in cgs units): 5.5 $\lesssim$ log $g$
$\lesssim$ 8 and  75000 K $\lesssim$
\teff $\lesssim$ 200000 K; see Werner \& Herwig
(2006) for a review. These hot stars are thought to be formed via a born--again
episode resulting either from a very late thermal pulse (VLTP) experienced by a
hot WD during its early cooling phase --- see Sch\"onberner
(1979) and Iben et al.  (1983) for earlier references --- or a late thermal
pulse (LTP) that occurs during the post-AGB evolution when hydrogen burning
is still active; see Bl{\" o}cker (2001) for references.  During the
VLTP, the helium flash driven  convection zone  reaches the hydrogen-rich 
envelope of the star, with the
consequence that most of the hydrogen content is burnt.  The star is
then forced to evolve rapidly back to the AGB and finally into the
domain of the PG1159 stars at high \teff\ values. LTP
also leads to a hydrogen--deficient composition but as a result of a
dilution episode. 

Interest in PG1159 stars is additionally motivated by the fact that eleven
of them exhibit multiperiodic luminosity variations
induced by nonradial $g-$mode pulsations. Pulsating PG1159 --- commonly
referred to as GW Virginis variables --- show low
degree ($\ell  \leq 2$), high  radial order ($k \gtrsim  18$) $g-$modes
with periods in the range from about 300 to 3000
s.  About half of them are still surrounded by a planetary nebula. 
Since the pioneering works of Starrfield et
al.  (1983, 1984) --- see also Starrfield et al. (1985) and Stanghellini et
al. (1991) --- the pulsation driving mechanism and in particular
the chemical composition of the driving zone in PG1159  stars have 
been the subject of hot debate. The works of Saio (1996)  and Gautschy
(1997), and more recently  Quirion et al.  (2004), Gautschy et al. (2005), 
and C\'orsico et al. (2006) have convincingly demonstrated
that $g$-mode pulsations in the range of periods of GW Virginis stars can
be  easily   driven  in   PG1159  models with a {\it uniform envelope 
composition} --- compatible with observed  photospheric  abundances  
 --- through the  $\kappa$-mechanism
associated  with  the  cyclical  partial  ionization  of  the  K--shell
electrons of  carbon and/or oxygen  in the envelope of these models.

A distinctive  feature characterizing  PG1159 stars is  their peculiar
surface  chemical composition.  Indeed, PG1159  stars  exhibit surface
layers   dominated   by   helium,   carbon,  and   oxygen,   the   main
nucleosynthetic constituents of the previous intershell layer of their
AGB progenitor stars. Typically, surface abundances\footnote{Throughout
this paper, we shall use abundances by mass fraction.} of about 0.33
He, 0.5 C,  and 0.17 O are reported, with  notable variations from star
to star (Dreizler  \& Heber 1998; Werner 2001).  In fact, almost every
star has its  individual He/C/O mixture. As reported  by Werner (2001)
and Werner  \& Herwig (2006)  the helium abundance ranges between  0.30--0.85,
and carbon  and oxygen abundances span  the ranges 0.15--0.60  and 0.02--0.20, 
respectively. In particular,  the helium
abundance covers mostly the range 0.3--0.5; only a minority of PG1159 stars
show a helium--enriched abundance in the range 0.6--0.85.

The dispersion in the atmospheric composition of PG1159 stars bears 
relevance in the question of the excitation of pulsation modes
in models of these stars, and particularly the coexistence of
variable and non-variable stars in the domain of the GW Virginis, a 
longstanding issue recently addressed by 
Quirion et al. (2004). The helium--enriched and 
nitrogen--deficient nonpulsating 
PG1159 star \mct\ is particularly outstanding. In fact, this
object --- characterized by a surface composition of about 0.74
He, 0.22 C,  and 0.03 O (Werner \& Herwig 2006) and a stellar mass of 
0.54 \msun\ (Miller Bertolami \& Althaus 2006) --- is well inside
the PG1159 instability domain in the \gteff\ diagram (C\'orsico et al.
2006). Even more remarkable is the fact that, in  
the \gteff\ diagram,  this star is located between two
pulsating stars, PG 2131$+$066 and PG 1707$+$427, either of them having 
standard
helium abundances (0.44). Other relevant
examples of nonpulsating PG1159 stars within the instability domain 
are the low--mass stars  \object{HS 0704$+$6153} with a surface helium abundance 
of 0.69 and  \hs1517 with a unsually high helium abundance of 0.85.
With the help of static PG1159 models, Quirion et al. (2004) have found that 
the high helium abundance would be
the cause for the absence of pulsations in \mct. Indeed, the envelope
of \mct\ is so helium--enriched as compared with its two neighbouring GW 
Virginis stars, that the ``helium poisoning'' would be responsible for the 
lack of pulsations in this star. However, it should be noted that the
no detection of pulsations in this object could be due to unfavorable
geometry. For instance if the star is observed pole-on,
modes with $m=\pm\ell$ would not be detected.

\begin{figure*}
\centering
\includegraphics[clip,width=350pt ]{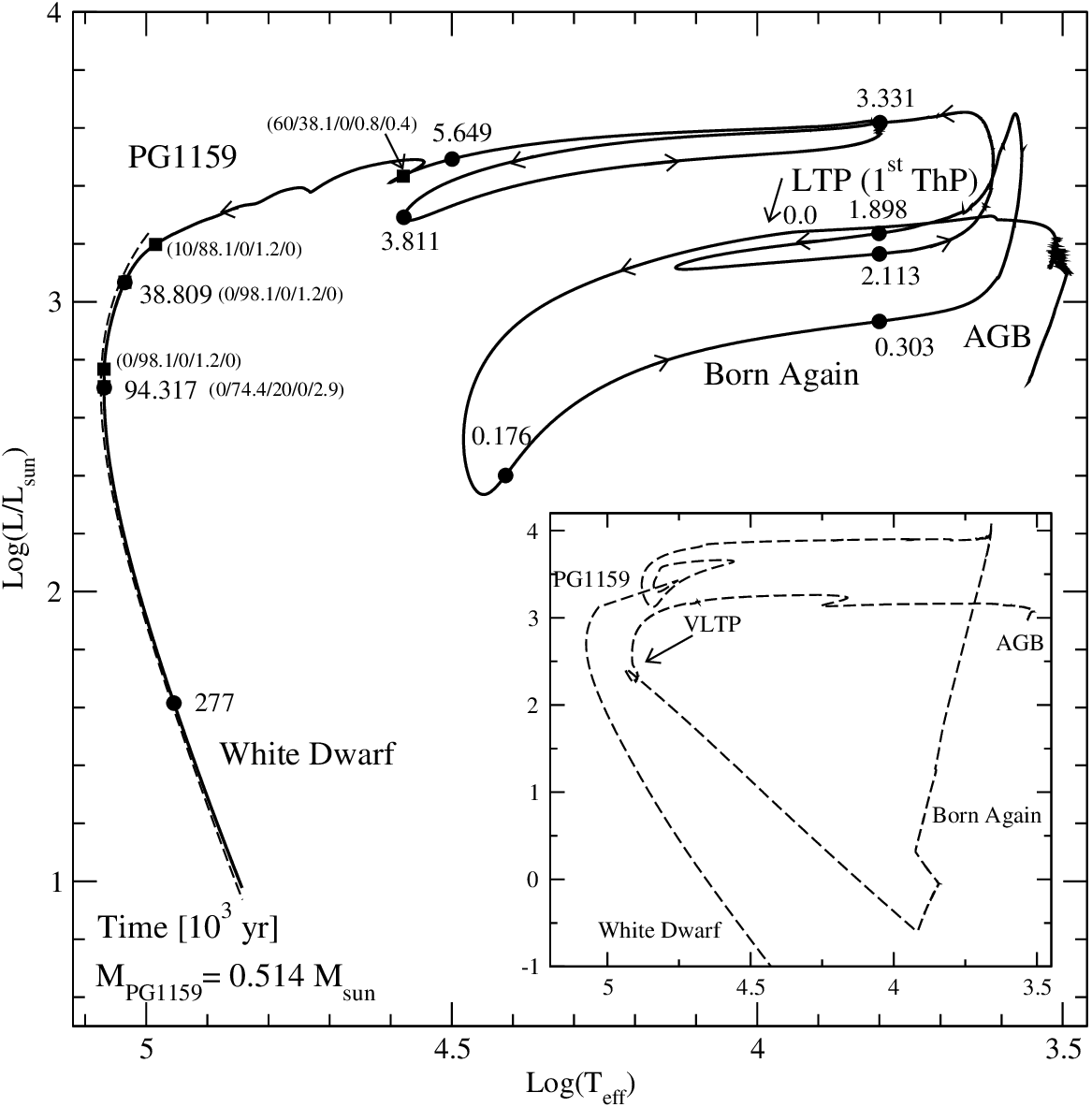}  
\caption {Hertzsprung-Russell diagram for the evolutionary scenario that 
leads to the formation of helium--enriched PG1159 stars.  The 
star experiences its first thermal pulse as a LTP. Numbers besides
filled circles and squares along the track give, respectively, the age 
(in $10^3$ yr) counted from the occurrence of the LTP and the surface
abundances ($^{1}$H/$^{4}$He/$^{12}$C/$^{14}$N/$^{16}$O) in percent. After the LTP, 
the remnant
star undergoes two main additional helium sub-flashes with the
consequent excursion towards lower temperatures. Note the changes in 
surface abundances as a result of mass--loss episodes. The helium--enriched 
intershell layer is uncovered well before evolution
reaches the domain of \mct\ at the onset of the hot WD stage. The evolutionary
track resulting from a VLTP episode is also shown with dashed lines for the
PG1159 regime; the complete track is depicted in the inset.}
\label{fig:hr}
\end{figure*}

The wide  variety of surface  patterns observed in PG1159  stars poses 
a major challenge to the theory of post-AGB evolution. Although
the dispersion in the surface composition observed in PG1159 stars can
admittedly  be understood in  terms of  the standard  born--again AGB
star scenario\footnote{The spread  in surface composition  of most
PG1159 stars can  be explained by differences in  stellar mass and the
number of thermal pulses  experienced by progenitors during the AGB.} (Herwig et al.  1999; Miller Bertolami \& Althaus  2006), the existence
of helium--enriched and nitrogen--deficient objects such as \mct\ is  
certainly more difficult
to explain and it still deserves  to be investigated.  It is precisely in
this vein that the present work is focused. Specifically,  we examine
the evolutionary scenario for the formation of  low--mass, helium--enriched  
PG1159 stars recently  proposed by Miller  Bertolami \&  Althaus (2006)
by assessing the overstability of pulsation $g-$modes of 
the resulting
stellar  models as  evolution  proceeds along  the  PG1159 domain.   
As  shown in  Miller  Bertolami \&
Althaus (2006)  this scenario could provide a possible  explanation for the
existence of PG1159 stars  such as the above mentioned 
\object{HS 0704$+$6153}, 
\object{HS  1517$+$7403}, and \mct,
three helium--enriched and nitrogen--deficient  objects in the sample of
Dreizler  \& Heber  (1998).   The following  section contains  details
about the main  aspects of this evolutionary scenario  and the stellar
models  we  used  as  a  background  for the pulsation stability  
analysis  described  in
Sect. 3.  Our  aim is to assess to what an extent the pertinent  
sequence of
stellar models do not show any sign of variability along the domain of
the GW Virginis stars, thus having a consistent picture of the
evolutionary and  pulsational status of \mct.
Finally, Sect. 4 is devoted to discussing and summarizing our results.
In particular, we discuss there the evolutionary connection 
\object{K1$-$27}\ $\rightarrow$ \hs1517\  $\rightarrow$ \mct\ that would link
the helium--rich O(He) stars with the helium--enriched PG1159 stars
and would explain the unusually high helium abundance observed in \hs1517.

\section{Evolutionary scenario}

The evolutionary scenario proposed by Miller Bertolami \& Althaus (2006)
involves the computation of
the advanced stages of the evolution of a low--mass star that, as a result
of mass--loss events, avoids the thermally pulsing AGB phase
and it experiences its first thermal pulse as a LTP after leaving the AGB.
Indeed,  helium
abundances  as high  as  observed in  the  helium--enriched PG1159 stars are
typical of the intershell helium abundances that develops during the 
{\it first} thermal pulse ($\sim$ 0.75\footnote{During  
subsequent  thermal pulses,  the
helium  abundance  in  the  intershell layer  gradually  decreases  if
convective  extramixing like  overshooting is  considered;  see Herwig
(2000). }).  
Because  the small  envelope  mass  characterizing
low--mass stars, this scenario appears to be quite likely in these stars.

\begin{figure}
\centering
\includegraphics[clip,width=242pt]{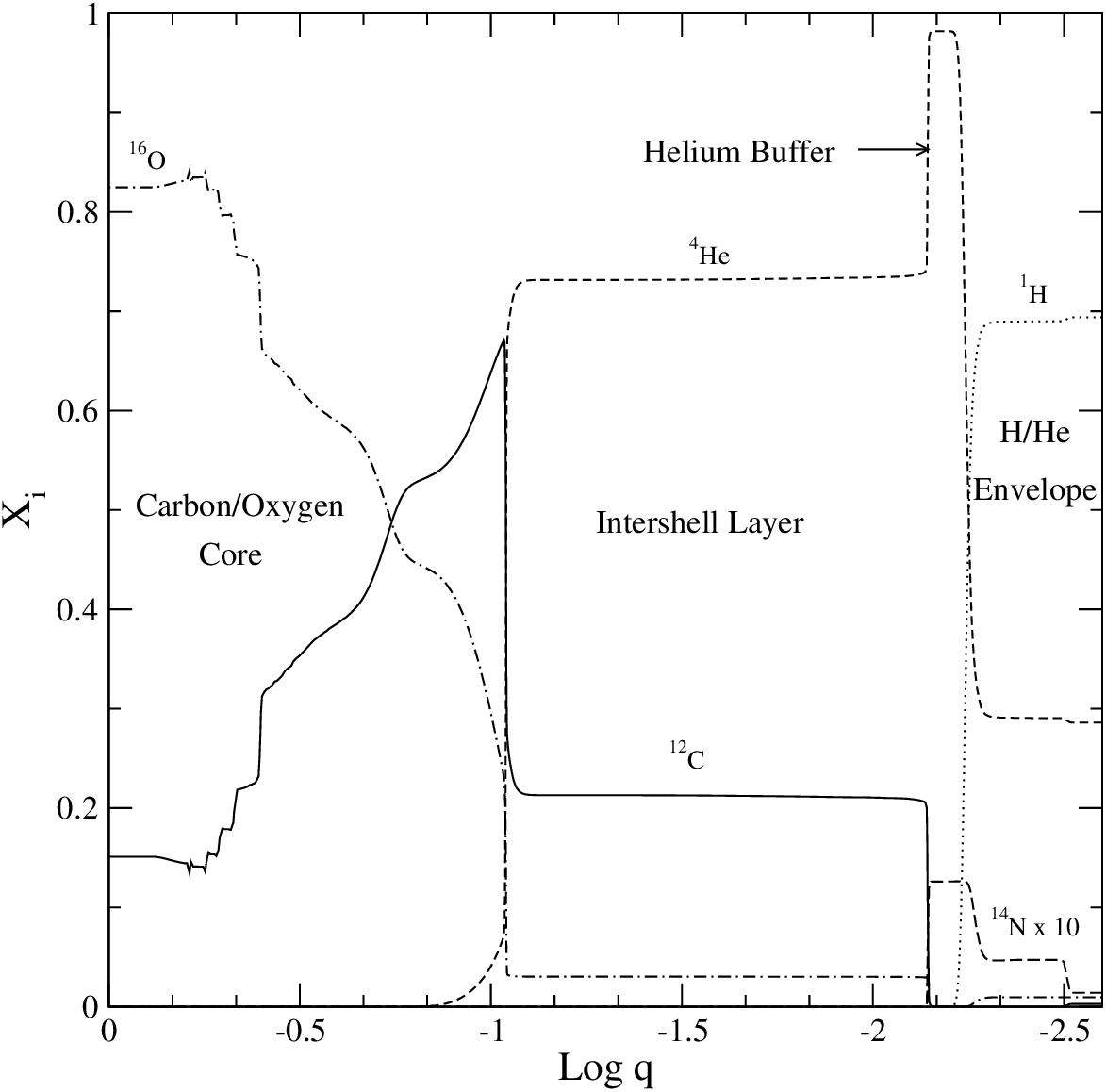}
\caption{Internal abundance distribution of  $^{1}$H, $^{4}$He, $^{12}$C,
$^{14}$N, and $^{16}$O as a function of the outer mass fraction $q$ 
for a model at 300 yr after the occurrence of the LTP (see Fig. 1). The
location of the helium buffer and the underlying intershell layer
are indicated.}
\label{fig:qui.eps}
\end{figure}

\begin{figure}
\centering
\includegraphics[clip,width=250pt]{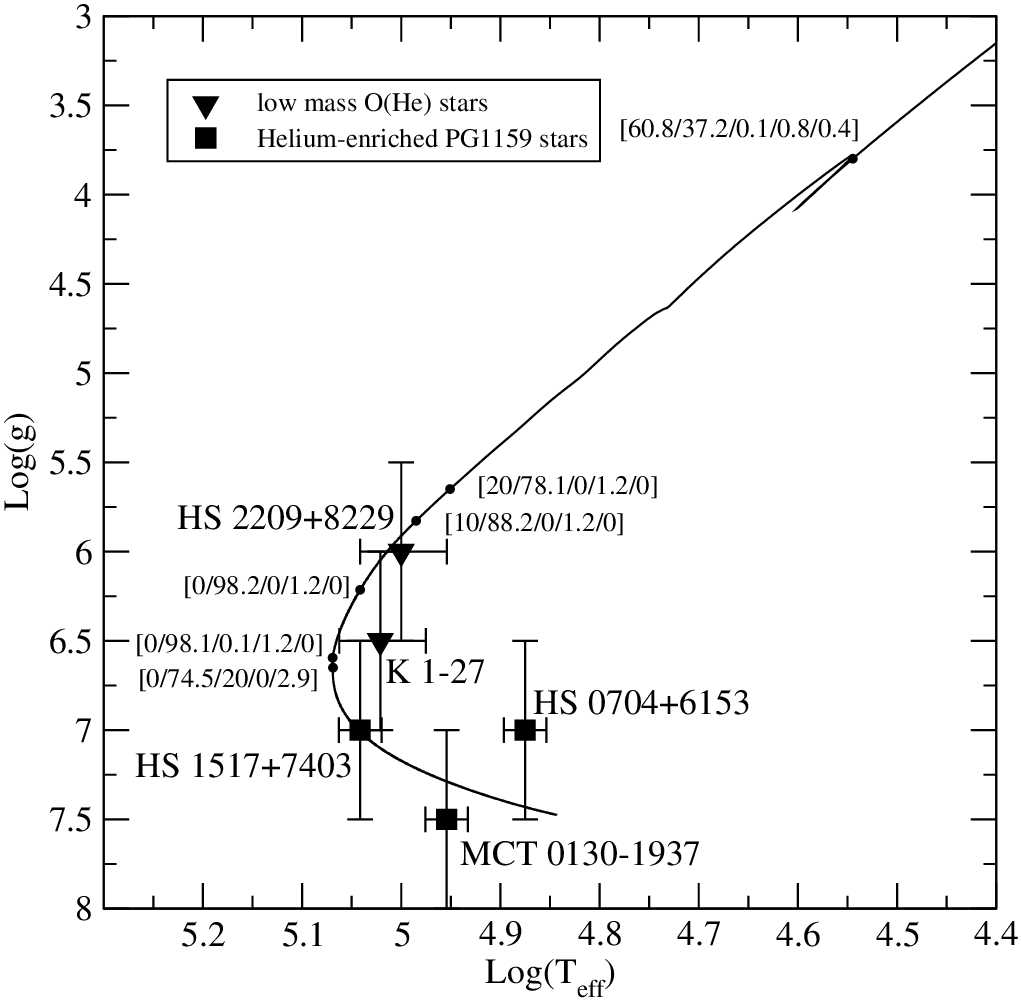}
\caption{\gteff\ evolution of the 0.514 \msun\ remnant. 
Surface abundances [$^{1}$H/$^{4}$He/$^{12}$C/$^{14}$N/$^{16}$O]  are given at 
selected stages (black dots). We also show the observational predictions 
for two low--mass O(He) stars and three  helium--enriched PG1159 stars 
(triangles and square symbols, respectively) taken from
Rauch et al. (1998) and Dreizler \& Heber (1998).}
\label{fig:gteff}
\end{figure}


The  Hertzsprung-Russell  diagram  corresponding to  the  evolutionary
scenario  is  shown  in  Fig. \ref{fig:hr}.  
Artificial mass--loss  rates force the initially  1-\msun\ model star
to undergo its first thermal pulse as a LTP at about 10000  K,
after leaving the AGB.  After the 
short-lived born--again episode and before the PG1159 domain, the  remnant 
experiences two additional excursions to lower temperatures --- the 
two double loop paths in  Fig.\ref{fig:hr} ---  by virtue of  helium  
sub-flashes. The total time  spent by the  remnant in the red during
these loop episodes amounts to about 3000 yr. During this time, 
mass--loss episodes are expected to erode the outer envelope 
considerably. For instance, mass--loss rates  
as high  as $10^{-5}$ \msun  /yr and  even larger  have been
observed  in Sakurai's  object (Hajduk et al. 2005).


In the interests of helping to understand the  evolutionary connections
to be studied later, we show in Fig. \ref{fig:qui.eps} the  chemical
stratification by  the end of the  born--again (about 300  yr after the
occurrence of the  LTP). This figure illustrates the inner
$^{1}$H,  $^{4}$He, $^{12}$C, $^{14}$N,  and $^{16}$O  distribution 
in terms  of the outer mass fraction $q$,
where $q=1-m_r/M_*$. 
Note the  presence of  the intershell  layer below  the thin
helium buffer, that results from the short-lived mixing episode during
the helium flash  at the LTP. This intershell layer  of 0.04 \msun\ is
substantially enriched  in helium and deficient  in nitrogen: [$^{4}$He,
$^{12}$C, $^{16}$O]=  [0.73, 0.21,  0.03]. Despite the  small envelope
(3.7  $\times  10^{-3}$ \msun)  overlying  the  intershell region,  no
strong dredge-up occurs. This is expected because of 
the low intensity of the first thermal pulse in low--mass stars.
By assuming mass--loss rates 
within observational expectations, Miller Bertolami \& Althaus (2006)
found that   the hydrogen--rich outer envelope  is eroded and surface 
abundances  start to change
gradually by log \teff\ $\sim$ 4.6, well before the  sequence reaches
the  domain of  the PG1159  stars.
From  this point until  the 
intershell chemistry  is uncovered, the nitrogen surface abundance
remains 0.012 with no  trace of carbon. 30000  yr later --- at  \teff\ $\sim$
108000 K --- the last vestiges of hydrogen--rich material left in the star
are removed and the surface exhibits the buffer abundances: [$^{4}$He,
$^{12}$C,$^{14}$N, $^{16}$O]=  [0.98, 0, 0.012, 0]. These  will be the
surface abundances until  the helium buffer is eroded by further
mass loss  during the  subsequent 50000 yr  of evolution, see 
Fig. \ref{fig:hr}.  During this
 time, the surface abundance will  be typical of
those  exhibited  by O(He)  stars.   Finally,  when  the helium buffer  is
removed    the    models    diplay    the    intershell    abundances ---
[$^{4}$He,$^{12}$C,$^{14}$N,   $^{16}$O]=  [0.74,  0.20,   0,  0.029] ---
characteristic of the  helium--enriched PG1159 stars.

The location of these
abundance changes in the \gteff\ diagram is shown in  Fig. \ref{fig:gteff}.
We also depict  the   location  of two
helium--rich O(He)  stars: \object{K1$-$27} and 
\object{HS 2209+8229}.  In particular \object{K1$-$27},
for which  detailed abundances have  been derived, displays 
an almost pure helium composition (0.98) with traces of $^{14}$N (0.017) --- 
however small hydrogen and carbon contents cannot be excluded, see Rauch 
et al. (1998) --- which are similar 
to the surface abundances predicted by the evolutionary scenario by the time 
the small  helium buffer is 
uncovered: 0.98 $^{4}$He and 0.012 $^{14}$N. In   addition,  we  include in 
the figure  the  three
helium--enriched   and   nitrogen--deficient   PG1159   stars  from  the
sample of non-pulsating stars of Dreizler  \& Heber  (1998): 
\object{HS  0704$+$6153}, \object{HS
1517$+$7403},  and \mct\   with  abundances  of  He/C/O=  0.69/0.21/0.09,
0.84/0.13/0.02,  and  0.74/0.22/0.03, respectively.   
Note  that the
evolutionary sequence  nearly reproduces both the location  in the \gteff\ 
diagram  and   the   surface  composition   of  the   three
helium--enriched PG1159 stars (in particular \mct ) and the O(He) 
star \object{K1$-$27}.  It is apparent from  Fig. \ref{fig:gteff} that
this  scenario suggests
an  evolutionary connection  between  the O(He)  stars  with low  mass
(namely \object{K1$-$27}  and \object{HS 2209+8229}) and the helium--enriched  PG1159 stars.
Note from Fig. \ref{fig:gteff} that  \object{K1$-$27} 
is consistent with a stellar mass value
of $M<$ 0.514 \msun. This is markedly lower than the value quoted
by Rauch et al. (1998) of 0.55 \msun. The difference can be traced back
to the fact that the Rauch et al. (1998) determination is based on the
Bl\"ocker (1995) hydrogen-burning evolutionary tracks, whilst our determination
is based on evolutionary sequences with surface abundances consistent with
those observed in this star\footnote{We are performing new simulations
that show that tracks for H and He burning post--AGB stars are remarkably
different for low--mass remnants ($M<0.55$ \msun).}. The same trend is also present in the new mass
determinations of PG1159 stars given in Miller Bertolami \& Althaus (2006),
as compared with the mass values quoted by Werner \& Herwig (2006), 
based also on hydrogen-burning post-AGB tracks.  

An issue of relevance as given by the evolutionary scenario concerns the
age predicted for \object{K1$-$27}. Specifically, we find that our sequence predicts
an age of about 30$-$60 Kyr at the observed effective temperature and
surface composition of \object{K1$-$27}. This age is a factor of 5$-$3 smaller than
the age quoted by Rauch et al. (1994), strongly alleviating the discrepancy
between the evolutionary age and the expansion age of the nebula\footnote{
We note that
the kinematical age of the nebula is not expected to change appreciably
as a result of our lower stellar mass for \object{K1$-$27}.}. This fact makes
the plausibility of the evolutionary scenario for the origin of this star 
even more attractive.

Needless to say, the exact  location in the \gteff\ diagram where 
the star will  show O(He)-  or helium--enriched  surface abundances
depends on the actual course  of mass--loss events. 
To  assess this,  we have re--computed  the post VLT  evolution of
the sequence but  for the case in which mass--loss  
rates are one  order of  magnitude lower  than assumed in Miller Bertolami
\& Althaus (2006).
In particular, we assume a  mass--loss rate during the giant state
of only $10^{-7}$ \msun /yr. As a result, the surface layers of the models
stay hydrogen-- and helium--rich throughout the evolution.  This
places a  lower limit --- below  observational expectations --- to  the 
mass--loss  rates  for which  the  evolutionary  scenario  may explain  the
existence of helium--enriched PG1159 stars like \mct.

Finally, we  note that the low--mass helium--enriched PG1159  stars could
be the result of a VLTP instead a  LTP.
To assess this we forced our post--AGB sequence to undergo a VLTP episode. In  
Fig. \ref{fig:hr} we include the resulting track. The full track is shown
in the inset.  Note that for the PG1159 regime, both tracks are almost
indistinguishable. 
As in  the case for the LTP, the  surface chemistry of the
emerging star  will be essentially that of  the intershell layer [$^{4}$He,
$^{12}$C, $^{16}$O]=  [0.74, 0.16, 0.03],  
but --- because of
the burning of the hydrogen content --- enriched in $^{13}$C and
nitrogen (0.01).  The  presence  of  nitrogen  is at  odds  with
observations  in the low--mass, helium--enriched  PG1159 stars. 
This fact discards the possibilty that the progenitors
of these stars have experienced a VLTP episode. Note also that because the 
entire helium--rich buffer is engulfed by the helium flash convection zone 
during the VLTP, the development of a helium--rich surface composition 
characteristic of the O(He) stars is certainly not expected in this case.
In addition, we want to comment on the fact  that from our numerical 
experiments the occurrence
of a VLTP episode in very low-mass stars ($M \lesssim 0.53$ \msun) appears to
require a delicate fine tuning in the mass--loss rate. Thus, we are
tempted to conclude that in these stars, VLTP episodes might be less 
likely than the LTP ones.

\begin{figure}
\centering
\includegraphics[clip,width=240pt]{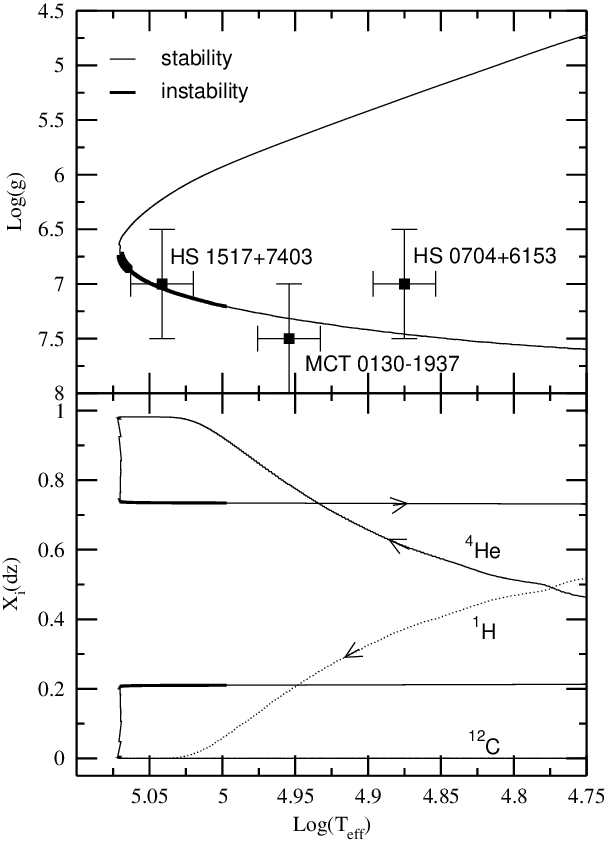}
\caption{Upper panel: \gteff\ evolution of our  0.514-\msun\ remnant 
and the location of the three helium--enriched PG1159 stars  according
to Dreizler \& Heber (1998):
\hs1517, \object{HS 0704$+$6153}, and \mct. The thick portion of the curve
corresponds to the instability domain of our 
sequence, and the thickest portion near the knee of the track
corresponds to the case for which the helium abundance of our 
model sequence
is artificially changed to 0.85 (see section 4). Bottom panel: 
Chemical abundances at the driving region (log $T \sim$ 6.2) in terms 
of the effective 
temperature. Arrows indicate the course of evolution.}
\label{track-abun}
\end{figure}

\begin{figure}
\centering
\includegraphics[clip,width=240pt]{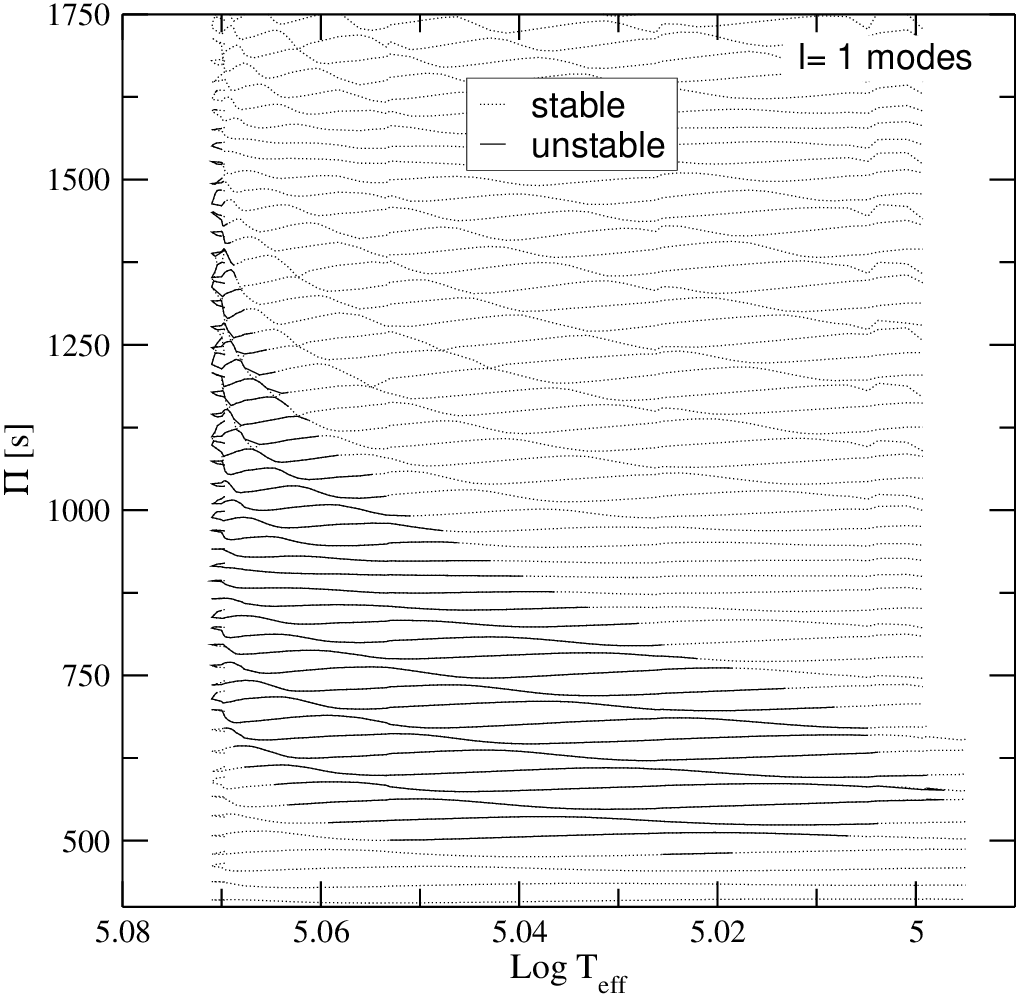}
\caption{Dipole modal diagram (periods vs. effective temperature) for our 
helium--enriched evolutionary sequence.
Solid (dotted) lines denote pulsational instability (stability). 
Unstable mode periods range from about 500 to 1600 s.
Below 
\teff\ $\sim$ 100000 K, no unstable pulsation modes are found.}
\label{zonas.eps}
\end{figure}

\section{Pulsational stability properties}

The evolutionary scenario proposed in Miller Bertolami \& Althaus (2006)
provides a possible explanation for the existence of helium--enriched
PG1159 stars like \mct, linking them with the low--mass O(He) stars.
In the frame of recent stability calculations (Quirion et al.
2004), we should expect that --- as a result of helium poisoning --- 
the pertinent stellar configurations do not show any 
sign of variability along the domain of the pulsating PG1159 stars. To 
assess this, we have performed pulsation stability analysis on our models.
It is important to note that we are examining the stability
properties of stellar models belonging to a real evolutionary
sequence derived from the complete history of the progenitor star, 
an aspect which renders robustness to our pulsational results.

We performed a pulsation  $g$-mode stability analysis by employing the
linear, nonradial, nonadiabatic  pulsation code  described in  
C\'orsico et  al.  (2006),
which  has  recently  been   employed  to  reexamine  the  theoretical
instability domain of pulsating PG1159 stars. This code assumes the 
``frozen-in convection''  approximation in which the perturbation
of    the    convective    luminosity    is   ignored.   


We  analyze the  stability  properties of stellar  models
covering  a range  of $5.1  \gtrsim \log
(T_{\rm eff})  \gtrsim 4.7$.  For  each model we have  restricted our
study to $\ell=  1$ $g$-modes with periods in the range  $ 50\ {\rm s}
\lesssim  \Pi \lesssim 5000$  s, thus  comfortably embracing  the full
period  spectrum  observed in  pulsating  PG1159 stars.   Surprisingly
enough, we find that, despite the high helium abundance in the driving 
layers, there exists  a region in the \gteff\ diagram  for  which our  
helium--enriched  PG1159  sequence
exhibits  unstable pulsation  modes. These modes are driven  by  
the $\kappa$-mechanism
associated with the  opacity bump due to partial  ionization of carbon
and oxygen; see Gautschy et al. (2005) and C\'orsico et al. (2006). 
To illustrate this,  we present in Fig.
\ref{track-abun} two panels which display the \gteff\ plane (upper  panel) and the chemical  abundances at the
driving region (bottom panel)  in terms of the effective  temperature.
Note the  presence of  a well-defined,
though short, instability  domain on the \gteff\
plane (thick  portion of the track).   It  is  clear that  the
extension  of the  instability  domain is  markedly  dependent on  the
helium abundance in the outer layers.   Indeed, it is only by the time
mass  loss  has eroded  the  helium  buffer and exposed the  helium--
and carbon--rich intershell layer --- bottom panel 
in Fig. \ref{track-abun} --- that our models start to 
exhibit pulsational instability.  From the modal diagram in 
Fig. \ref{zonas.eps}, this instability starts to manifest itself at
 the longest pulsation periods. Note that unstable mode
periods range from about 500 to 1600 s. Note also the strong  decrease  in 
the  longest expected periods; this is because the
high helium content  in the  driving  region. By the time  evolution 
has proceed to \teff\ $\sim$ 100000 K our models cannot drive unstable  
modes anymore and they become stable with further evolution (see 
Fig. \ref{zonas.eps}). This behaviour 
is in sharp contrast with the situation encountered in low--mass PG1159 models
with standard surface helium abundances, which  show
pulsational instability throughout this region of the 
\gteff\ diagram (C\'orsico et al. 2006). 

It is clear from Fig. \ref{track-abun} that the 
nonpulsating \mct\ star is located well  outside the theoretical 
instability domain.
In fact, we do not find any sign of instability in our models 
by the time evolution has reached the domain of \mct. This result hints
at a consistent picture between the evolutionary scenario described in 
Section 2 that could
explain the existence of helium--enriched PG1159 stars and the 
nonvariable nature of \mct.  


\section{Discussion and conclusions}

We have examined the evolutionary scenario proposed by Miller Bertolami 
\& Althaus (2006) that could  explain the  existence of low--mass, 
helium--enriched PG1159 stars. 
In particular, this scenario remarkably reproduces both  
the location  in  the \gteff\
diagram and  the helium--enriched and nitrogen--deficient composition
of \mct, and suggests as well a possible  evolutionary  connection   
between  the low--mass helium--rich O(He) stars (namely  \object{K1$-$27} and 
\object{HS 2209+8229}) and  the helium--enriched  PG1159 stars.


In this paper we have assessed the overstability of $g-$modes of the pertinent
stellar models as evolution proceeds  along the PG1159 domain. We find
that --- despite the  high helium abundance in the  driving layers --- there
exists a region in the \gteff\ diagram for which
our helium--enriched PG1159  sequence exhibits unstable pulsation modes
with  periods  in  the  range  500   to  1600  s.  This is a novel
aspect contained in  this study. The domain of  instability is
restricted to a rather narrow region of the \gteff\  diagram. 
In  particular, \mct\ is located outside the theoretical 
instability domain.
In this  sense, this finding reinforces the conclusions arrived  at in
Quirion et al. (2004) about  the nonvariability of \mct.  This result hints
at a  consistent picture  between  the evolutionary  scenario proposed
by Miller Bertolami \& Althaus (2006) for the origin of low--mass,  
helium--enriched PG1159  
stars  and  the nonvariable nature of \mct.   We conclude 
that \mct\ is likely a {\it real} non-pulsating star and that the
lack of pulsations cannot be attributed to unfavorable geometry.

However, as documented in Fig. \ref{track-abun}, the nonpulsating 
helium--enriched object \hs1517 
lies well inside the predicted instability domain of our sequence.
The presence of this non-variable star in the 
unstable region of our sequence could be understood in terms of its higher
surface helium abundance (about 0.85, see Werner \& Herwig 2006) 
as compared with that of \mct. In fact, the helium
surface abundance of \hs1517 is intermediate 
between the intershell helium abundance (0.75) of our models and that of the
helium buffer (0.98) for which no unstable modes are found at all.
For a quantitative inference, we have recomputed the post born--again 
evolution of our
sequence by artificially changing the  surface abundance to be compatible
with that of \hs1517. Our pulsational analysis shows in this case
that, though a very narrow
instability domain still persists --- see Fig. \ref{track-abun} --- 
\hs1517 is outside  
the resulting theoretical unstable region.  A helium abundance as high as 
that observed in \hs1517 appears to be required to starve pulsations in
this star.  In contrast, for a helium  abundance of 0.75 we saw 
that unstable modes are excited. This theoretical finding strongly supports
the observational expectation for the unusually high helium abundance
in \hs1517.  

From an evolutionary point of view, the high helium abundance
in  \hs1517 is difficult to understand.
Indeed, helium abundances larger than about 
0.75 are not expected in the intershell layer during the AGB evolution. 
In addition, the presence of 
abundant carbon at its surface --- about 0.13 by mass, Werner \& Herwig (2006) 
--- reflects the occurrence of helium burning in prior evolutionary stages. 
It is conceivable that the surface composition observed
in \hs1517 could be reflecting the chemical abundance distribution 
existing in the narrow transition layer between the helium buffer
and the massive intershell region --- see Fig. \ref{fig:qui.eps}. 
This layer comprises only about 6.5 $\times  10^{-5}$ \msun. With our adopted
mass--loss rates, this layer is rapidly eroded in a matter of 6500 yr. During
this time  our sequence barely evolves in  the \gteff\ diagram 
--- see Fig. \ref{fig:gteff}. But for mass--loss 
rates similar to those charaterizing O(He) stars (about $10^{-9}$ \msun /yr) --- a strong reduction in the mass--loss rate
is indeed expected with decreasing luminosity --- this layer would be
eroded slow enough 
for the remnant to evolve to the domain of \hs1517 with surface
abundances similar to those observed in this star. This prompts us to suggest
that  \hs1517 could be a transition object between the low--mass O(He) 
stars and the helium--enriched PG1159 stars like \mct.

The evolutionary scenario proposed by Miller Bertolami \& Althaus (2006) 
suggests the possibility that low--mass O(He)  stars could be the direct 
progenitors
of the  helium--enriched PG1159  stars, and not form  a distinct 
post--AGB evolutionary channel --- for instance the result of a stellar 
merging event.
The plausibility of this scenario is sustained not only by spectroscopic
evidence but also,
as shown in this work, by consistent pulsational stability calculations of  
helium--enriched PG1159 models.  In addition, the substantially
smaller ages predicted by this scenario for \object{K1$-$27} are in line with
the expected kinematical age of the nebula. Finally, 
we have  put forward the possibility of an evolutionary connection 
\object{K1$-$27}\  $\rightarrow$ \hs1517\  $\rightarrow$ \mct. 
The existence of 
these evolutionary links appears more attractive in view of the 
observational fact that the helium--enriched PG1159  stars
are nitrogen--deficient, which rules out the occurrence of a VLTP episode 
during the progenitor evolution. 
In case that a VLTP had occurred, then an homogeneous composition 
--- corresponding to the intershell chemistry --- throughout
the envelope would have been expected immediately after the born again, 
with the consequent result that no marked surface abundance change 
would have resulted from mass--loss episodes during the further evolution.
Finally, detailed tabulations of the calculations presented here are
available at our web site: http://www.fcaglp.unlp.edu.ar/evolgroup/.

\begin{acknowledgements}

We warmly acknowledge the comments and suggestions of our referee, which
strongly improve both the contents and presentation of the paper.
This research was supported by the Instituto de Astrof\'{\i}sica La
Plata and by the PIP 6521 grant from CONICET. We thank H. Viturro 
and R. Martinez for technical support.


\end{acknowledgements}

\end{document}